\documentclass[onecolumn,showpacs,preprintnumbers,amsmath,amssymb]{revtex4}
\usepackage{amssymb,amsmath,amsthm,mathtools}
\usepackage{hyperref}
\usepackage{url,graphicx,epstopdf} 
\usepackage{graphicx}
\usepackage{dcolumn}
\usepackage{bm}
\usepackage{epsfig}
\usepackage{subfigure}
\usepackage{xcolor}
\usepackage{bbold}

\newcommand{\bi}{\begin{itemize}}
\newcommand{\ei}{\end{itemize}}
\newcommand{\be}{\begin{eqnarray}}
\newcommand{\ee}{\end{eqnarray}}
\newcommand{\beq}{\begin{equation}}
\newcommand{\eeq}{\end{equation}}
\newcommand{\bbmatrix}{\left( \begin{array}}
\newcommand{\eematrix}{\end{array} \right)} 

\def\ch#1{\textcolor{black}{#1}}

\begin{document} 

\title{Optimal Control for Quantum Metrology via Pontryagin's principle}

\author{Chungwei Lin$^1$\footnote{clin@merl.com}, Yanting Ma$^1$, Dries Sels$^{2,3,4}$}

\affiliation{$^1$Mitsubishi Electric Research Laboratories, 201 Broadway, Cambridge, MA 02139, USA \\ 
$^2$Department of physics, Harvard University, Cambridge, MA 02138, USA \\ 
$^3$Department of physics, New York University, New York City, NY 10003, USA \\
$^4$Center for Computational Quantum Physics, Flatiron Institute, 162 5th Ave, New York, NY 10010, USA
}

\date{\today}

\begin{abstract} 
Quantum metrology comprises  a set of techniques and protocols that utilize quantum features for parameter estimation which can in principle outperform any procedure based on classical physics.  
We formulate the quantum metrology in terms of an optimal control problem and apply Pontryagin's Maximum Principle to determine the optimal protocol that maximizes the quantum Fisher information for a given evolution time. 
As the quantum Fisher information involves a derivative with respect to the  parameter which one wants to estimate, we devise an augmented dynamical system that explicitly includes gradients of the quantum Fisher information. The necessary conditions derived from Pontryagin's Maximum Principle are used to quantify the quality of the numerical solution. 
The proposed formalism is generalized to problems with control constraints, and can also be used to maximize the classical Fisher information for a chosen measurement.
\end{abstract}

\maketitle 

\section{Introduction}  
Modern quantum technology \cite{Neilson_book, book:Kaye_book, Shor:1997:PAP:264393.264406,  RevModPhys.66.481, RevModPhys.77.513, LIGO_2011} requires manipulating the wave function to achieve performance beyond the scope of classical physics.
A typical quantum task starts from an easily prepared initial state, undergoes a designed control protocol, and hopefully ends up with a state sufficiently close to the target state. When the closeness to the target state can be quantified by a scalar metric (a terminal cost function), 
the quantum task can be formulated as an optimal control problem -- one tries to find the best control protocol that maximizes the performance metric for given resources. Many quantum applications (or at least an intermediate step of the application) fit this description. Important examples include quantum state preparation \cite{PhysRevA.97.062343,  ahmed19, PhysRevA.74.022306, PhysRevX.8.021012, PhysRevLett.112.047601, PhysRevA.90.013409, PhysRevA.93.013851} where the cost function is the overlap to the known target state, the ``continuous-time'' variation-principle based quantum computation \cite{Farhi_00, PhysRevLett.103.080502, PhysRevA.90.052317, Peruzzo-2014, PhysRevX.6.031007} where the cost function is the ground state energy, and quantum parameter estimation (quantum metrology) \cite{book:Helstrom, book:Holevo, PhysRevLett.96.010401, Brif_2010, Glaser-2015, AdvancesQuantumMetrology_2011, PhysRevX.6.031033, PhysRevA.96.040304,  RevModPhys.90.035005, PhysRevX.10.031003, PhysRevLett.110.053002, Sekatski2017quantummetrology, PhysRevA.96.032310, doi:10.1116/5.0006785, PhysRevLett.125.020402} where the cost function is the Fisher information.  
Maximal Fisher information has been used for optimal estimation of Hamiltonian parameters  \cite{Pang-2017,  PhysRevLett.123.260505, PhysRevLett.124.060402, Koczor_2020}. Numerically, the Fisher information can be optimized by e.g. GRAPE (GRadient Ascent Pulse Engineering \cite{Khaneja-2005}) both for single and multiple parameter estimations in the presence of noise \cite{PhysRevA.96.012117, PhysRevA.96.042114, PhysRevResearch.2.033396, Liu_2019}. 
The Fisher information has also been used to quantify the precision to which certain parameters of external signals (external to the sensing qubit) can be measured \cite{PhysRevA.82.012337, PhysRevX.8.021059, Muller-2018}. For quantum metrology application, optimal control has been applied to the preparation of entangled superposition states that are required for optimal measurement, e.g., squeezed spin states \cite{PhysRevLett.100.250406, PhysRevA.93.013851} or Ramsey interferometry with BEC on atom-chips \cite{PhysRevA.93.010304, va-2014}.

Pontryagin's Maximum Principle (PMP) \cite{book:Pontryagin, Sussmann_87_01, book:Luenberger, book:GeometricOptimalControl} is a powerful tool in classical control theory, and it has been applied to quantum state preparation \cite{PhysRevA.97.062343, PhysRevA.101.022320} and non-adiabatic quantum computation \cite{PhysRevX.7.021027, PhysRevA.100.022327}. In essence, PMP adopts the variational principle to derive a set of necessary conditions for the optimal control. In particular, it provides an efficient way to compute the gradient of the cost function with respect to the control field as well as the evolution time by introducing an auxiliary system (described by costate variables) that follows the dynamics similar to the original problem. When the system degrees of freedom are small (such as a single qubit), these necessary conditions are very restrictive and analytical solutions can sometimes be constructed \cite{PhysRevA.97.062343, PhysRevA.100.022327, PhysRevA.101.022320}. For systems of higher dimensions, these necessary conditions become less informative but the efficient procedure of computing gradient is still useful for numerical solutions. Moreover, PMP optimality conditions are valuable in quantifying the quality of a numerical solution and can be done with almost no extra computational overhead. In this work, we extend PMP to quantum metrology applications where the natural choice of the terminal cost function is the quantum/classical Fisher information (QFI/CFI). 
The fact that QFI/CFI involves a derivative with respect to the external parameter causes some non-trivial complications. To properly use PMP, we devise an augmented dynamical system that involves the variables appearing in QFI, based on which the switching functions can be stably and efficiently obtained. With the provided formalism, we are able to numerically demonstrate that the optimal control indeed satisfies all the necessary conditions of PMP. 

The rest of the paper is organized as follows. First we describe the concrete problem and review the necessary background of PMP. We then introduce the augmented dynamics that is designed for QFI and CFI. The formalism will be applied to a few problems, including maximizing QFI within a given control constraint and maximizing CFI for a given measurement basis. A short conclusion is provided in the end.

\section{Problem and augmented dynamics for PMP}
The concrete problem we consider is the ``twist and turn'' Hamiltonian \cite{RevModPhys.90.035005, Hayes_2018, PhysRevLett.124.060402}
\begin{equation}
H(t) = \chi \hat{J}_z^2 + \omega \hat{J}_z + \Omega(t) \hat{J}_x,
\label{eqn:tt_Hamil}
\end{equation} 
\ch{with $[\hat{J}_i, \hat{J}_j] = i\, \epsilon_{ijk} \hat{J}_k$ ($i=x,y,z$) } and the initial state the non-entangled maximum-eigenvalue state of $\hat{J}_x$, denoted as $| \Psi_\text{coh-x} \rangle$. The potential physical realizations include  interacting (generalized) spins \cite{PhysRevLett.123.260505, PhysRevX.10.031003}, the two-arm interferometer \cite{PhysRevA.85.022322, PhysRevA.33.4033}, and superradiance \cite{PhysRev.93.99, GROSS1982301}. 
The goal of the control is to efficiently estimate the parameter $\omega$ (around zero) in Eq.~\eqref{eqn:tt_Hamil}, i.e., to produce a final state $| \psi(T) \rangle$ over the total evolution time $T$ that is as sensitive as possible to the change of the parameter $\omega$ around zero. The quantitative metric is QFI: 
\begin{equation}
\mathcal{F}_Q( |\psi(T) \rangle) = 4\left[ 
\langle \partial_\omega \psi(T) | \partial_\omega \psi(T) \rangle -  |\langle  \psi(T) | \partial_\omega \psi(T) \rangle|^2 \right].
\end{equation} 
In Eq.~\eqref{eqn:tt_Hamil}, $\hat{J}_z^2$ is the source of entanglement and referred to as a ``twist'' term; $\hat{J}_x$ is the external control and referred to as a ``turn'' term. For  eigenstates of $\hat{J}_z$, denoted as $|m\rangle_z$, $\hat{J}_z^2$ determines their relative phases but not amplitudes whereas $\hat{J}_x$ determines their relative amplitudes but not phases. 
The optimal control problem is to find an $\Omega(t)$ that steers $|\psi(0) \rangle = | \Psi_\text{coh-x} \rangle$ to a final state $|\psi(T) \rangle$ that maximizes QFI at a given terminal time $T$. Using the terminology of control theory, Eq.~\eqref{eqn:tt_Hamil} is control-affine as it depends linearly on the control $\Omega(t)$, and is time-invariant as \ch{ the time dependence of $H(t)$ is exclusively through the control $\Omega(t)$}. 

\ch{Hamiltonian~\eqref{eqn:tt_Hamil} represents a set of $N$ all-to-all interacting spins where $\hat{J}_i = \sum_{n=1}^N \frac{\sigma_i}{2}$ ($i=x,y,z$ and $\sigma$'s are Pauli matrices)}.
For a system composed of $N$ spins, QFI$(t)$ = $Nt^2$ is referred to as the ``shot-noise'' limit (SNL) which can be achieved without any quantum entanglement; QFI$(t)$ = $N^2 t^2$ as the Heisenberg's limit (HL) which is the upper bound of QFI and is achieved by preparing the initial state as $| \Psi_\text{HL} \rangle = ( |M\rangle_z + |-M\rangle_z )/\sqrt{2}$ with $|\pm M\rangle_z$ the largest/smallest-eigenvalue eigenstate of $\hat{J}_z$ (the maximum eigenvalue $M=N/2$) \cite{PhysRevLett.96.010401}. A system displays quantum enhancement when QFI is larger than SNL. 
One of the key insights from Haine and Hope in Ref.\cite{PhysRevLett.124.060402} is that for a limited evolution time $T$, the process of state preparation (i.e., to produce an entangled state) should also be regarded as a degree of freedom to maximize QFI. This becomes essential when $T$ is too short (small $N \chi T$) to produce a highly entangled state.  


To compute QFI we need $| \partial_\omega \psi(T) \rangle$ which can be obtained by evolving $| \partial_\omega \psi(t) \rangle$ via the differential equation  $
\partial_t | \partial_\omega \psi(t) \rangle 
= -i \hat{J}_z | \psi(t) \rangle 
- i H(\omega, t) | \partial_\omega \psi(t) \rangle
$ and the initial condition $ | \partial_\omega \psi(t=0) \rangle = 0$. To apply PMP, we regard $| \psi \rangle$ and $| \partial_\omega \psi \rangle$ as {\em independent} dynamical variables. 
Denoting $| \psi \rangle$ as $| \psi_0 \rangle$, $| \partial_\omega \psi \rangle$ as $| \psi_1 \rangle$, the {\em augmented} dynamics satisfies
\beq 
\begin{aligned}
\partial_t \begin{bmatrix} | \psi_0 \rangle \\ | \psi_1 \rangle\end{bmatrix} = 
\begin{bmatrix} -i H(\omega) & 0 \\ 
-i \hat{J}_z  & -i H(\omega) \end{bmatrix} 
\begin{bmatrix} | \psi_0 \rangle \\ | \psi_1 \rangle \end{bmatrix}.
\end{aligned}
\label{eqn:dyna_2c}
\eeq 
The initial augmented state is  $(| \psi_0 \rangle, | \psi_1 \rangle) = (| \Psi_\text{coh-x} \rangle, 0)$. The terminal cost function (to minimize)  is 
\beq 
\mathcal{C}_\text{Q} = - ( \langle \psi_1(T) | \psi_1(T) \rangle - | \langle \psi_1 (T) | \psi_0(T) \rangle |^2 \rangle ),
\label{enq:cost_2c}
\eeq  
which, up to a positive factor, is the {\em negative} QFI. The subscript 'Q' indicates the quantum case.  

Given a dynamical system Eq.~\eqref{eqn:dyna_2c}, PMP introduces a set of auxiliary costate variables based on which the switching function and control Hamiltonian (c-Hamiltonian) are defined. Following the standard procedure \cite{PhysRevA.97.062343, PhysRevA.100.022327, PhysRevA.101.022320, book:Luenberger}, we denote $| \pi_0 \rangle$ and $| \pi_1 \rangle$ as the costate variables (in the form of wave function) of $| \psi_0 \rangle$ and $| \psi_1 \rangle$, and derive their dynamics as
\beq 
\partial_t \begin{bmatrix} | \pi_0 \rangle \\ | \pi_1 \rangle\end{bmatrix} = 
\begin{bmatrix} -i H(\omega) & -i \hat{J}_z \\ 
0  & -i H(\omega) \end{bmatrix} 
\begin{bmatrix} | \pi_0 \rangle \\ | \pi_1 \rangle \end{bmatrix},
\label{eqn:dyna_pi_2c}
\eeq 
with the costate boundary conditions 
\beq 
\begin{aligned}
|\pi_0(T) \rangle &= \frac{\partial \mathcal{C}_\text{Q} }{\partial \langle \psi_0(T) |} =  +  | \psi_1 (T) \rangle  \langle \psi_1 (T) | \psi_0 (T) \rangle, \\
|\pi_1(T) \rangle &= \frac{\partial \mathcal{C}_\text{Q} }{\partial \langle \psi_1(T) |} =  -| \psi_1(T) \rangle +  | \psi_0 (T) \rangle  \langle \psi_0 (T) | \psi_1(T) \rangle.
\end{aligned}
\label{eqn:pi_boundary}
\eeq 
Note that $\langle \psi_0 (t) | \psi_1 (t) \rangle = 0$ when $\omega=0$. 
The switching function $\Phi(t)$ and c-Hamiltonian $\mathcal{H}_c(t)$ are  
\beq 
\begin{aligned}
\Phi(t) &= \text{Im} \left\{
\begin{bmatrix} \langle \pi_0 | & \langle \pi_1 | \end{bmatrix} 
\begin{bmatrix} \hat{J}_x & 0 \\ 0 & \hat{J}_x \end{bmatrix}
\begin{bmatrix} | \psi_0 \rangle \\ | \psi_1 \rangle \end{bmatrix} 
\right\}, 
\\ 
\mathcal{H}_c(t) &= \text{Im} \left\{
\begin{bmatrix} \langle \pi_0 | & \langle \pi_1 | \end{bmatrix} 
\begin{bmatrix} H(\omega) & 0 \\ \hat{J}_z & H(\omega) \end{bmatrix}
\begin{bmatrix} | \psi_0 \rangle \\ | \psi_1 \rangle \end{bmatrix} 
\right\}.
\end{aligned}
\label{eqn:Phi_Hc_Coupled}
\eeq 
According to PMP, $\Phi(t) \sim \frac{\delta \mathcal{C} }{\delta \Omega(t) }$ and  $\mathcal{H}_c \sim \frac{\partial \mathcal{C} }{\partial T}$. The necessary conditions for an optimal control $\Omega(t)$ are (i) $\Phi(t)=0$ and (ii) $\mathcal{H}_c(t)$ is a constant over the entire evolution time $T$ \cite{book:Luenberger, book:GeometricOptimalControl}. \ch{Condition (i) is general (optimal solution requires a zero gradient with respect to the cost function) whereas condition (ii) is specific to time-invariant control problems; both can be served to quantify the control quality.} Practically, $\Phi(t)$ can be used in the gradient-based optimization algorithm (i.e., $\Omega(t) \rightarrow \Omega(t) -\gamma \Phi(t)$ with a learning rate $\gamma$) for numerical solutions. 
The sign of $\mathcal{H}_c(t)$ tells if increasing the evolution time $T$ reduces ($\mathcal{H}_c(t)<0$) or increases ($\mathcal{H}_c(t)>0$) the terminal cost function \cite{PhysRevA.101.022320}. In all our simulations, $\mathcal{H}_c(t)<0$ meaning increasing the evolution time increases QFI. This holds for unitary dynamics but is not expected to be the case in the presence of quantum decoherence.

Three general remarks are pointed out. First, as the dynamics based on Schr\"odinger equation [Eq.~\eqref{eqn:tt_Hamil}] is typically control-affine, the optimal control is expected to contain some ``bang'' sector(s) \cite{PhysRevX.7.021027, PhysRevA.97.062343,  PhysRevA.100.022327, PhysRevA.101.022320}. Based on the control theory, this expectation requires a terminal cost function that is also linear in $| \psi(T) \rangle$, which is true when using the fidelity as the terminal cost for a known target state \cite{PhysRevA.97.062343,  PhysRevA.100.022327, PhysRevA.101.022320}. As QFI is quadratic in the final state, the optimal control is not expected to be bang-bang in general. 
Second, the augmented dynamics \eqref{eqn:dyna_2c} is non-unitary. This is not essential for the formalism but imposes demands on the numerical ODE (ordinary differential equation) solver. In the implementation we express the dynamics using real-valued variables and use the explicit Runge-Kutta method of order 5 as the ODE solver. 
Finally, the proposed formalism regards $|\psi \rangle$, $| \partial_\omega \psi \rangle$ as {\em independent} dynamical variables and introduces $|\lambda_0 \rangle$, $|\lambda_1 \rangle$ as their corresponding costate variables. \ch{ Compared to the GRAPE algorithm  where computing the gradient $\frac{\delta \mathcal{C} }{\delta \Omega(t) }$ at each $t$ requires an integration over time (Appendix of Ref.~\cite{PhysRevA.96.012117}), in the proposed formalism the gradient $\frac{\delta \mathcal{C} }{\delta \Omega(t) } \sim \Phi(t)$ is {\em local} in time [Eq.~\eqref{eqn:Phi_Hc_Coupled}], greatly reducing the  computation complexity. }
We notice that the forward augmented dynamics alone [Eq.~\eqref{eqn:dyna_2c}] can be used to compute the gradients with respect to multiple control parameters \cite{doi:10.1063/1.3267086} and has been applied to construct the optimal gate operations \cite{doi:10.1063/1.3267086, PhysRevLett.120.150401}. 
Before moving to concrete examples we point out that the model considered here [Eq.~\eqref{eqn:tt_Hamil}] contains three parameters: the number of spins $N$, the twist strength $\chi$, and the total evolution time $T$. In the following discussions $T=1$ unless assigned specifically. We now apply the formalism to analyze a few interesting cases.  

\section{Applications}

\subsection{Convergence of optimal control}   

\begin{figure}[ht]
\centering
\includegraphics[width=0.9\textwidth]{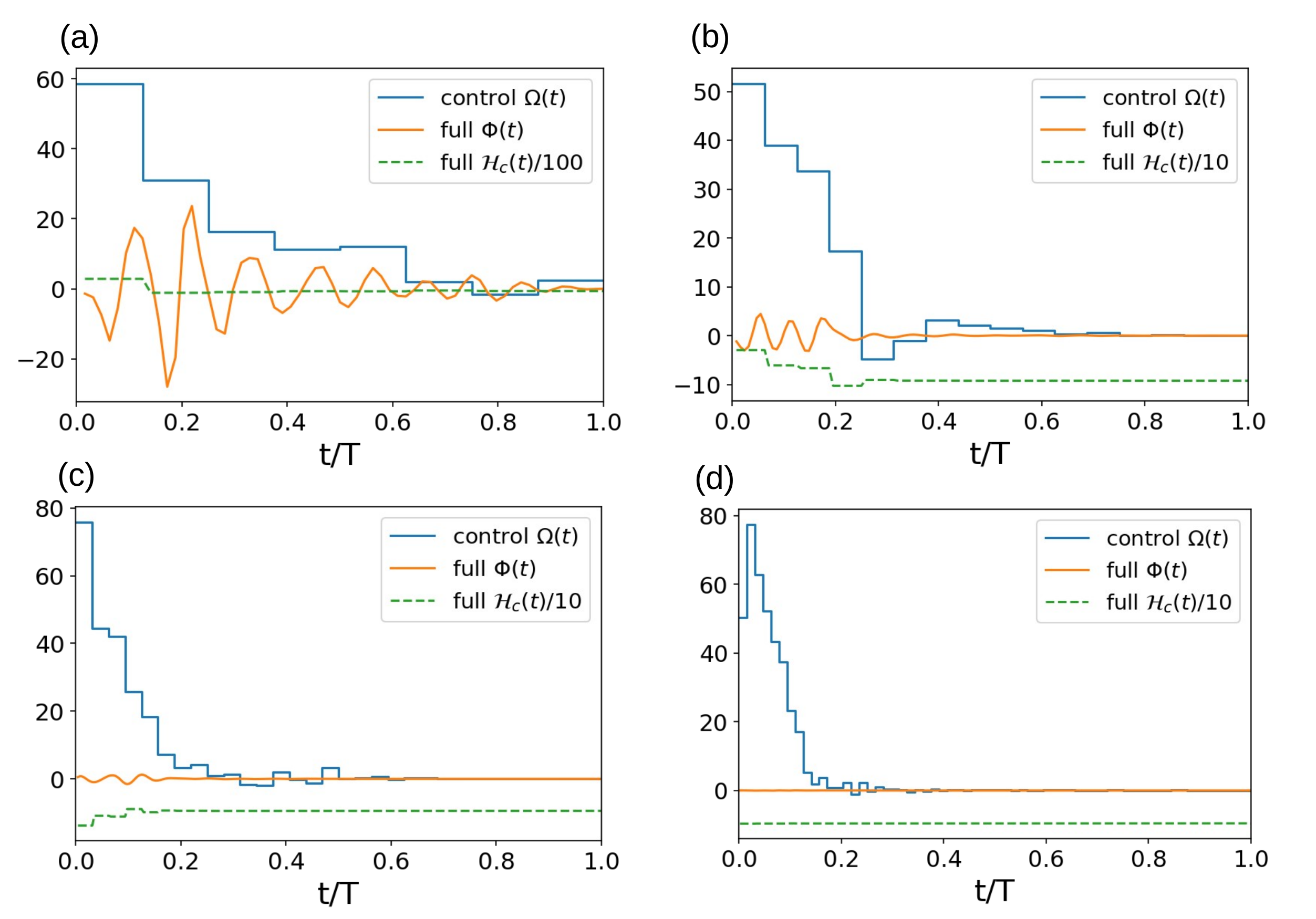}
\caption{ The optimal controls using 8 (a), 16 (b), 32 (c), and 64 (d) controls (time intervals) with $(N, \chi)=(20, 4)$.  Increasing the number of controls reduces the magnitudes of the switching function and results in a flatter negative c-Hamiltonian. 
}
\label{fig:N20_chi4_Tpts}
\end{figure} 


The control function $\Omega(t)$ is typically approximated by a piece-wise constant function, i.e., $\Omega(t) = \Omega_i$ for $t \in [t_i, t_{i+1})$, with the evolution time $T$ divided into $N_t$ equal time intervals \cite{Khaneja-2005}. As the first application, we investigate how the optimal QFI converges upon increasing $N_t$ to approximate $\Omega(t)$. The motivation is to quantify the solution quality from the smallness of the switching function, which can be characterized by a mean and a standard deviation: 
\beq 
\begin{aligned}
\Phi_\text{m} &\equiv \frac{1}{T} \int_0^T dt\, \Phi(t), \\
\Phi_\text{sd} &\equiv \frac{1}{\sqrt{T}} \left[ \int_0^T dt\, \Phi^2(t) \right]^{1/2}.
\end{aligned}
\label{eqn:Phi_mean_sd}
\eeq  
The normalizations are chosen such that $\Phi_\text{m}$ and $\Phi_\text{sd}$ have no dependence on $T$.
For an optimal control, $\Phi(t)=0$ so both $\Phi_\text{m}$ and $\Phi_\text{sd}$  vanish. When the piece-wise constant function is approaching to the optimal solution, $\Phi_\text{m}$ is also close to zero and the value of $\Phi_\text{sd}(>0)$ can be used to characterize how good a solution is. 

\begin{table}[ht]
\begin{tabular}{l | lll l }
$(N,\chi)$ & QFI$_{8}$/$\Phi_\text{sd}$ & QFI$_{16}$/$\Phi_\text{sd}$ & QFI$_{32}$/$\Phi_\text{sd}$ & QFI$_{64}$/$\Phi_\text{sd}$ 
\\ \hline
(10,4) &  80.16/8.96$\times 10^{-1}$  & 87.96/4.72$\times 10^{-2}$ &  88.15/3.00$\times 10^{-3}$ & 88.15/2.10$\times 10^{-3}$  \\
(20,1) &  270.13/6.62$\times 10^{-1}$  & 273.19/4.34$\times 10^{-2}$ & 273.28/6.10$\times 10^{-3}$ & 273.28/5.55$\times 10^{-3}$   \\ 
(20,2) & 320.38/1.72  &  330.26/3.00$\times 10^{-1}$ & 331.86/1.81$\times 10^{-2}$  &  331.88/1.00$\times 10^{-2}$ \\
(20,4) & 223.31/7.95 &  341.35/1.11 & 356.37/3.2$\times 10^{-1}$ & 364.60/1.08$\times 10^{-2}$   \\
(30,1) & 648.24/4.12   &   659.27/4.14$\times 10^{-1}$ &   661.74/4.04$\times 10^{-2}$ &  661.78/2.86$\times 10^{-2}$
\end{tabular}
\caption{Optimal (maximum) QFI for different number of controls. The subscript denotes the number of controls $N_t$; $\Phi_\text{sd}$ is defined in Eq.~\eqref{eqn:Phi_mean_sd}. $(N, \chi)$=(10,4), (20,1), (20,2), (20,4), (30,1) are considered. For a given $(N, \chi)$, a control resulting in a smaller standard deviation $\Phi_\text{sd}$ has a larger QFI. }
\label{table:QFI_Ncontrols}
\end{table} 

Table \ref{table:QFI_Ncontrols} summarizes the optimal QFI's for $(N, \chi)$=(10,4), (20,1), (20,2), (20,4), (30,1) using different number of controls and their corresponding $\Phi_\text{sd}$'s. As expected, the control that results in a smaller  $\Phi_\text{sd}$ gives a larger QFI. Fig.~\ref{fig:N20_chi4_Tpts} plots the optimal $\Omega(t)$ and the corresponding $\Phi(t)$ and $\mathcal{H}_c(t)$ for $(N, \chi)$=(20,4) using 8, 16, 32, 64 controls. The optimal control using  more time intervals gives a smaller switching function and a flatter (negative) c-Hamiltonian.

\begin{figure}[ht]
\centering
\includegraphics[width=0.9\textwidth]{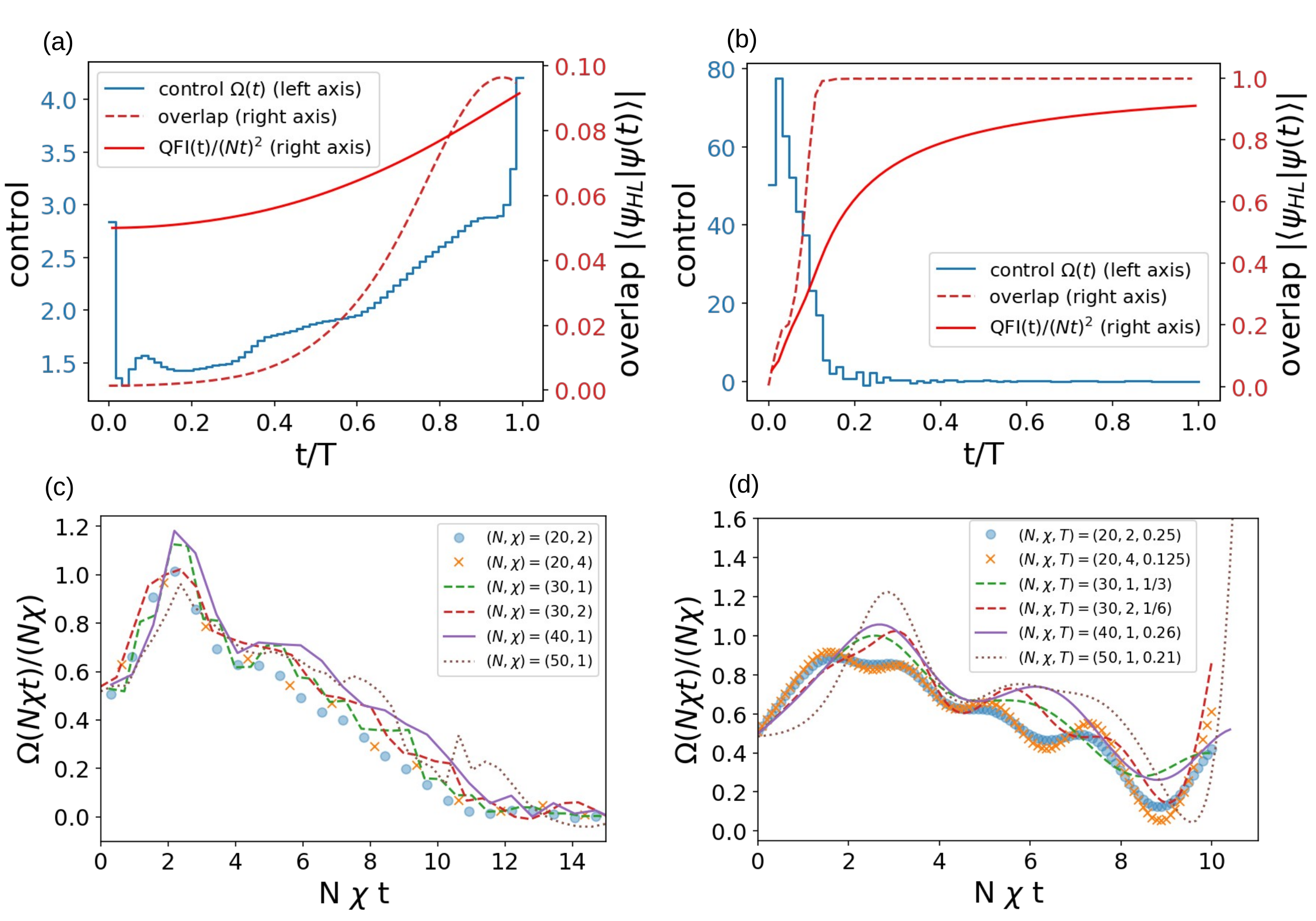}
\caption{ The optimal controls (blue, left axis), the overlap  $| \langle \psi_\text{HL}| \psi(t) \rangle |$ (red dashed, right axis), and QFI/$(Nt)^2$ (red solid, right axis) for ($N$, $\chi$) = (20, 0.1) (a) and (20,4) (b). In (a), $\Omega(t)$ is non zero during the whole evolution, and the overlap to $| \psi_\text{HL} \rangle$ increases to $\lesssim 0.1$ around $t=0.96$. In contrast in (b), the optimal control $\Omega(t)$ vanishes after $t \gtrsim 0.18$ around which $| \langle \Psi_\text{HL}| \psi(t) \rangle |$ is approaching one. 64 controls are used in these simulations. (c) The dimensionless controls [Eq.~\eqref{eqn:dimensionless}] for $(N, \chi) = $ (20,2), (20,4), (30,1), (30,2), (40,1) and (50,1): they almost collapse to a single curve. (d) The dimensionless controls using $|\Psi_\text{HL} \rangle$ as the target state for $(N, \chi, T) = $ (20,2,1/4), (20,4,1/8), (30,1,1/3), (30,2,1/6), (40,1,0.26) and (50,1,0.21). All overlaps $| \langle \Psi_\text{HL}| \psi(T) \rangle |$ are larger than 0.985.
} 
\label{fig:N20_Heisenberg_overlap}
\end{figure} 

\subsection{Strong twist limit}

When $N \chi$ is large, the optimal control appears to be strong during early evolution and vanish after a certain amount of time (the same observation is also pointed out in Ref.~\cite{PhysRevLett.124.060402}). This behavior can be understood by invoking the state that achieves HL. If preparing $| \psi_\text{HL} \rangle$ takes only a small fraction of the total evolution time $T$, one way to maximize QFI is to first produce $| \Psi_\text{HL} \rangle$ and then let system interact freely with the environment. The resulting QFI is roughly $N^2 (T-t_\text{prep})^2$ which approaches the HL $N^2 T^2$ when $T \gg t_\text{prep}$ ($t_\text{prep}$ is time to produce $| \psi_\text{HL} \rangle$).
To see what the optimal control does,  Fig.~\ref{fig:N20_Heisenberg_overlap}(a) and (b) contrast the optimal controls for $\chi=0.1$ and 4 using $N= 20$. For $\chi=0.1$, $\Omega(t)$ is non-zero over the entire $T$; for $\chi=4$, $\Omega(t)$ vanishes around $t=0.18$. The overlap $|\langle \psi_\text{HL}| \psi(t) \rangle|$ and the normalized QFI/$(Nt)^2$  are also provided. As $\hat{J}_x$ is the only term in Eq.~\eqref{eqn:tt_Hamil} capable of changing the $| m \rangle_z$ population, $\Omega(t)$ has to be non-zero to change the overlap; once $\Omega(t)$ is zero the value of $|\langle \psi_\text{HL}| \psi(t) \rangle|$ is fixed.  
For $\chi=0.1$ where the entanglement source is too weak to bring the state close to $| \psi_\text{HL} \rangle$, the control is always non-zero and $|\langle \psi_\text{HL}| \psi(t) \rangle|$ gradually increases to $\lesssim 0.1$. 
For $\chi=4$ where the entanglement source is strong, the control steers the state close to $| \psi_\text{HL} \rangle$ during $t \leq 0.18$ and then is turned off; QFI is maximized via steering the state to $| \psi_\text{HL} \rangle$ fast. This behavior appears to be general once $N \chi$ is sufficiently large [Fig.~\ref{fig:N20_Heisenberg_overlap}(c)].  


One can further analyze the the optimal control by 
expressing Eq.~\eqref{eqn:tt_Hamil} at $\omega=0$ as
\beq 
\begin{aligned}
&i \partial_t | \psi \rangle  = N \chi \left[ \frac{\hat{J}_z^2}{N} +\frac{\Omega(t)}{N \chi} \hat{J}_x \right] | \psi \rangle  \\ 
\Rightarrow & i \frac{\partial | \psi \rangle }{\partial (N \chi t) } = 
\left[ \frac{\hat{J}_z^2}{N} + \left( \frac{\Omega(t)}{N \chi} \right) \hat{J}_x \right] | \psi \rangle.
\end{aligned}
\label{eqn:Sch_2}
\eeq 
with the given initial state $| \Psi_\text{coh-x} \rangle$. The additional information in the strong $N \chi$ limit is that target state is also known, at least approximately, to be $| \Psi_\text{HL} \rangle$. The second expression of Eq.~\eqref{eqn:Sch_2} is scaled such that the spectral ranges of $\hat{J}_z^2/N$ and $\hat{J}_x$ are comparable for {\em all} $N$ and therefore $\Omega(t)/(N \chi)$ represents the $N$-independent strength ratio between the ``twist'' and ``turn''. Eq.~\eqref{eqn:Sch_2} also introduces a dimensionless time $N \chi t$. Denoting $\Omega_{N \chi}(t)$ to be the optimal control for a given $(N, \chi, T=1)$, we define the corresponding dimensionless control as 
\beq 
\bar{\Omega}_{N \chi} (t) \equiv \frac{ \Omega_{N \chi}(N \chi t) }{ N \chi }.
\label{eqn:dimensionless}
\eeq 
Because of similar structures of the initial and target states for all $N$ (i.e., $| \Psi_\text{coh-x} \rangle$  is peaked at $|m=0 \rangle_z$ and is monotonously decreased as $|m|$ increases; $| \Psi_\text{HL} \rangle$ is only non-zero at $|m=\pm N/2 \rangle_z$), the dimensionless control $\bar{\Omega}_{N \chi} (t)$ is expected to be only weakly dependent on $N \chi$.
Fig.~\ref{fig:N20_Heisenberg_overlap}(c) gives $\bar{\Omega}_{N \chi} (t)$ for $(N, \chi) = $ (20,2), (20,4), (30,1), (30,2), (40,1) and (50,1): their optimal dimensionless controls [Eq.~\eqref{eqn:dimensionless}] to a good approximation collapse to a single curve. A direct consequence is that the total input energy to maximize QFI, defined by $\int_0^T dt \,\Omega(t) $, is roughly a constant. 

To examine this picture more carefully, 
Fig.~\ref{fig:N20_Heisenberg_overlap}(d) shows the dimensionless controls that maximize $| \langle \Psi_\text{HL} | \psi(T) \rangle|^2$ \cite{PhysRevA.100.022327, PhysRevA.101.022320} for $(N, \chi, T) = $ (20,2,1/4), (20,4,1/8), (30,1,1/3), (30,2,1/6), (40,1,0.26) and (50,1,0.21). The evolution time $T$ is chosen to be close to and smaller than the optimal time  (i.e. a $T$ such that $\mathcal{H}_{c}$ being small and negative) and all overlaps $| \langle \Psi_\text{HL}| \psi(T) \rangle |$ are larger than 0.985; note the product $N \chi T \sim 10$. 
The optimal controls based on maximizing $| \langle \Psi_\text{HL} | \psi(T) \rangle|^2$  [Fig.~\ref{fig:N20_Heisenberg_overlap}(d)] share the following features: (i) $\Omega(t)$ has a peak around $N \chi$ at $t \sim T/3$; (ii) $\Omega(t)$ increases drastically from $\Omega \sim 0$ around $t \lesssim T$. Feature (i) is captured by the optimal controls that maximize QFI [Fig.~\ref{fig:N20_Heisenberg_overlap}(c)] but   (ii) is not because maximizing QFI requires turning off $\Omega$ once the state reaches $| \Psi_\text{HL} \rangle$. Feature (ii) thus highlights the difference between maximizing $| \langle \Psi_\text{HL} | \psi(T) \rangle|^2$  (the traditional sensing intuition that separates the state preparation and the state evolution \cite{PhysRevLett.124.060402}) and maximizing QFI in the strong $N \chi$ limit.

\subsection{System with constrained control amplitude}

\begin{figure}[ht]
\centering
\includegraphics[width=0.9\textwidth]{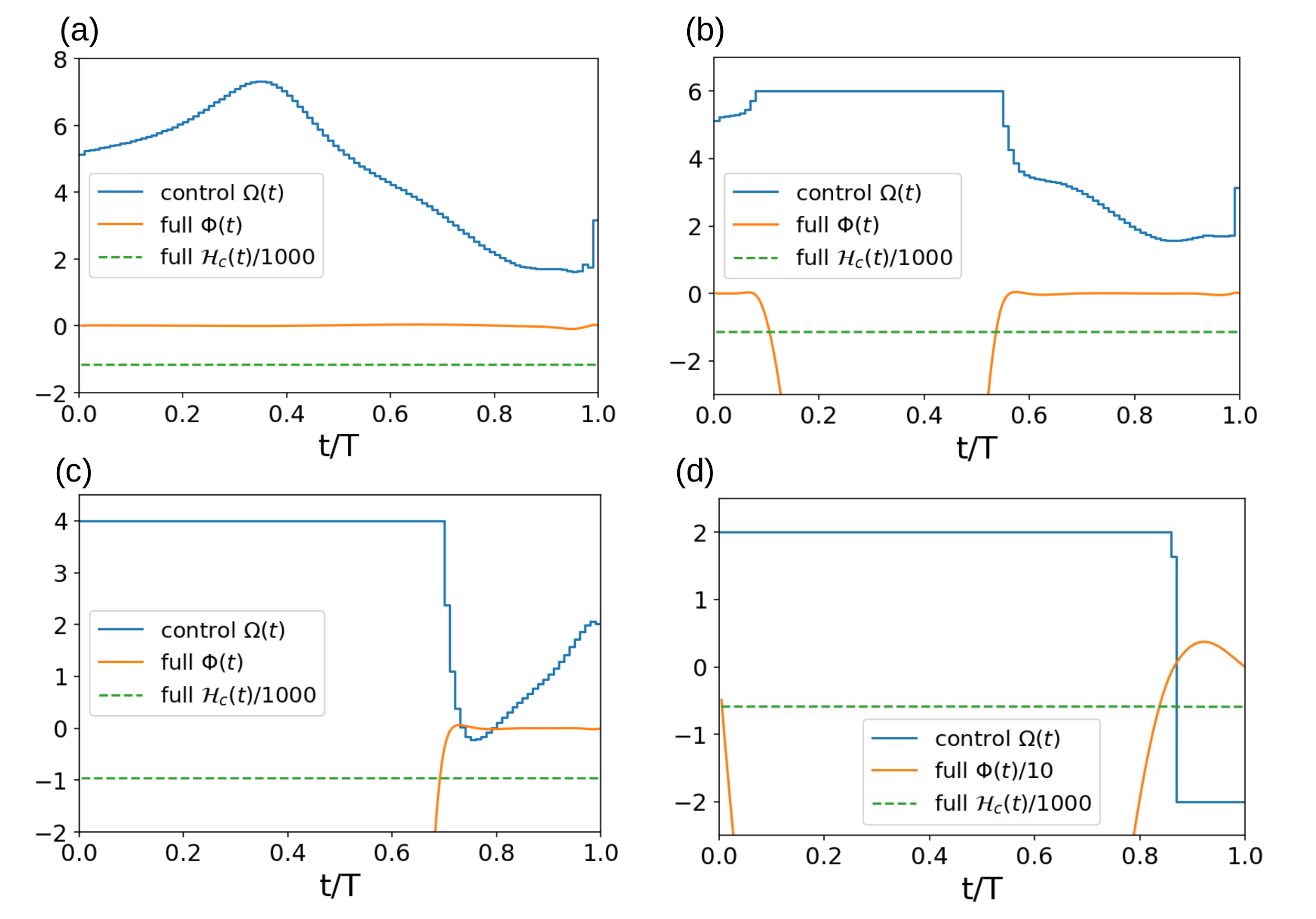}
\caption{Optimal control of $N=100$, $\chi=0.1$ using 100 time intervals. (a) No constraint on the control. (b)-(d) Optimal controls with $\Omega(t) <u_\text{max}$: (b) $u_\text{max}=6$; (c) $u_\text{max}=4$; (d) $u_\text{max}=2$.  When the optimal control takes one of the extreme values, $\Phi(t)$ has an opposite sign.
}
\label{fig:N100_chi0.1_constraint}
\end{figure}

As a second application, we consider  $N=100$, $\chi=0.1$. These parameters are used as an example in Ref.~\cite{PhysRevLett.124.060402}. With the ability to compute the gradient efficiently, we use 100 time intervals to approximate $\Omega(t)$ and the obtained optimal control is given in Fig.~\ref{fig:N100_chi0.1_constraint}(a). We see that the necessary conditions are to a good approximation satisfied; specifically $|\Phi_\text{m}| \lesssim 10^{-3}$ and $\Phi_\text{sd} \approx 0.006$.

In practice the control amplitude is bounded, i.e., $|\Omega(t)| \leq u_\text{max}$ and to obtain the optimal control with amplitude constraint requires an additional step during the iteration: $\Omega(t)$ is taken to be the closest extreme (bang) value when $|\Omega(t)|>u_\text{max}$. The necessary condition is modified: when $\Omega(t)$ takes the extreme value, the sign of the switching function $\Phi(t)$ is opposite to that of $\Omega(t)$; otherwise $\Phi(t)=0$. Fig.~\ref{fig:N100_chi0.1_constraint}(b)-(d) show the results of $u_\text{max} = 6, 4, 2$. We see that the necessary conditions are well satisfied. Imposing the maximum $|\Omega(t)|$ reduces the optimal QFI from 2895.0 (no constraint), 2869.9 ($u_\text{max}=6$), 2431.1 ($u_\text{max}=4$), to 1347.5 ($u_\text{max}=2$).  Consistent with the intuition, the bang control appears when $| \Omega^*(t) | > u_\text{max}$ with $\Omega^*(t)$ the optimal control without constraints.

\subsection{Classical Fisher Information}

As a final application, we use PMP to maximize CFI defined as 
\beq 
\mathcal{F}_C = \sum_m \frac{ (\partial_\omega {P}_m)^2 }{ {P}_m }.
\label{eqn:classical_FI}
\eeq  
$P_m = | _x\langle m | e^{i \phi \hat{J}_z} | \psi(T) \rangle  |^2$ is the probability distribution of $\hat{J}_x$ measurement ($| m \rangle_{x,z}$'s are eigenstates of $\hat{J}_{x,z}$). Following Ref.~\cite{PhysRevLett.119.193601, PhysRevLett.124.060402} an additional phase offset $\phi$ is introduced. 
The terminal cost function (to minimize) is chosen to be $\mathcal{C}_\text{C} = -\mathcal{F}_C$ (the subscript 'C' indicates ``classical''), and the most crucial step is to compute $ \frac{\partial \mathcal{C}_\text{C}}{\partial \langle \psi_{0,1} (T)|}$ to get the boundary condition for the costates $| \pi_{0,1}(T) \rangle$.

Denoting the solution of Eq.~\eqref{eqn:dyna_2c} at $t=T$ to be $| \psi_0(T) \rangle = \sum_m \bar{\alpha}_m | m \rangle_z $ and $| \psi_1(T) \rangle = \sum_m \bar{\beta}_m | m \rangle_z$, and applying $e^{i \hat{J}_z \phi}$ to the terminal state leads to 
\beq 
\begin{aligned}
e^{i \hat{J}_z \phi} | \psi_0(T) \rangle 
&= \sum_m e^{i \phi m} \bar{\alpha}_m | m \rangle_z 
= \sum_m \alpha_m | m \rangle_x = \sum_m 
\left[ \sum_n U_{mn} \bar{\alpha}_n e^{i n \phi} \right] | m \rangle_x, \\
e^{i \hat{J}_z \phi} | \psi_1(T) \rangle 
&=\sum_m e^{i \phi m}\bar{\beta}_m | m \rangle_z   =  \sum_m \beta_m | m \rangle_x = 
\sum_m  \left[ \sum_n U_{mn} \bar{\beta}_n e^{i n \phi} \right] | m \rangle_x.
\end{aligned}
\eeq 
What we have directly from Eq.~\eqref{eqn:dyna_2c} are $\bar{\alpha}_n$, $\bar{\beta}_n$, and $U_{mn}$ (where each row vector of $U$ is an eigenvector of $\hat{J}_x$ and $U$ is real-valued in z-basis), based on which we get $\alpha_m = \sum_{n'} U_{mn'} \bar{\alpha}_{n'} e^{+i n' \phi}$, $\beta_m = \sum_{n'} U_{mn'} \bar{\beta}_{n'} e^{+i n' \phi}$, $P_m = \alpha_m^* \alpha_m$, $\partial_\omega {P}_m = \beta^*_{m} \alpha_{m} + \alpha^*_{m} \beta_{m}$ and 
\beq 
\mathcal{F}_C =  \sum_m \frac{ (\beta^*_{m} \alpha_{m} + \alpha^*_{m} \beta_{m}  )^2}{ \alpha^*_{m} \alpha_{m} }.
\label{eqn:classical_FI_2}
\eeq  
Straightforward derivatives give   
\beq 
\begin{aligned}
\frac{ \partial \mathcal{F}_C }{ \partial \bar{\alpha}_n^*} &=
\sum_m \frac{ - (\partial_\omega {P}_m)^2 }{ {P}_m^2  } 
\frac{\partial {P}_m }{\partial \bar{\alpha}^*_n }
+  \sum_m \frac{ 2 (\partial_\omega {P}_m) }{ {P}_m  } \frac{\partial ( \partial_\omega {P}_m) }{\partial \bar{\alpha}^*_n }, \\
\frac{ \partial \mathcal{F}_C }{ \partial \bar{\beta}_n^*} &=
\sum_m \frac{ 2 (\partial_\omega {P}_m) }{ {P}_m  } \frac{\partial ( \partial_\omega {P}_m) }{\partial \bar{\beta}^*_n }, \\
\frac{ \partial \mathcal{F}_C }{ \partial \phi} &=
\sum_m \frac{ - (\partial_\omega {P}_m)^2 }{ {P}_m^2  } 
\frac{\partial {P}_m }{\partial \phi }
+  \sum_m \frac{ 2 (\partial_\omega {P}_m) }{ {P}_m  } \frac{\partial ( \partial_\omega {P}_m) }{\partial \phi }
\end{aligned}
\label{eqn:partial_Fc}
\eeq 
where $\frac{\partial {P}_m }{\partial \bar{\alpha}^*_n }
=  \left[ U_{mn} e^{-i n \phi}\right] \alpha_m$, $\frac{\partial ( \partial_\omega {P}_m) }{\partial \bar{\alpha}^*_n } = \left[ U_{mn} e^{-i n \phi}\right] \beta_m$, $ 
\frac{\partial ( \partial_\omega {P}_m) }{\partial \bar{\beta}^*_n } = \left[ U_{mn} e^{-i n \phi}\right] \alpha_m$, and
\beq 
\begin{aligned} 
\frac{\partial {P}_m }{\partial \phi } &= 
\left[\sum_n U_{mn} (-i \cdot n) e^{-i n \phi} \bar{\alpha}^*_n\right] \alpha_m + c.c., \\
\frac{\partial ( \partial_\omega {P}_m) }{\partial \phi} &= 
\left[\sum_n U_{mn} (-i \cdot n) e^{-i n \phi} \bar{\beta}^*_n\right] \alpha_m 
+ \beta^*_m \left[\sum_n U_{mn} (+i\cdot  n) e^{+i n \phi} \bar{\alpha}_n \right]
+c.c..
\end{aligned}
\eeq 
The negative of Eq.~\eqref{eqn:partial_Fc} is used as the terminal boundary condition of the costate variables $|\pi_0 (t) \rangle$, $|\pi_1 (t) \rangle$, i.e.,
\beq
\begin{aligned}
|\pi_0 (T) \rangle &= \sum_n \left[ -
\frac{ \partial \mathcal{F}_C }{ \partial \bar{\alpha}_n^*} \right] | n \rangle_z, \\
|\pi_1 (T) \rangle &= \sum_n \left[ -
\frac{ \partial \mathcal{F}_C }{ \partial \bar{\beta}_n^*} \right] | n \rangle_z.
\end{aligned} 
\label{eqn:pi_T_in_z}
\eeq 
The phase $\phi$ is updated by $\phi \rightarrow \phi + \gamma \frac{ \partial \mathcal{F}_C }{ \partial \phi}$. 


\begin{figure}[ht]
\centering
\includegraphics[width=0.9\textwidth]{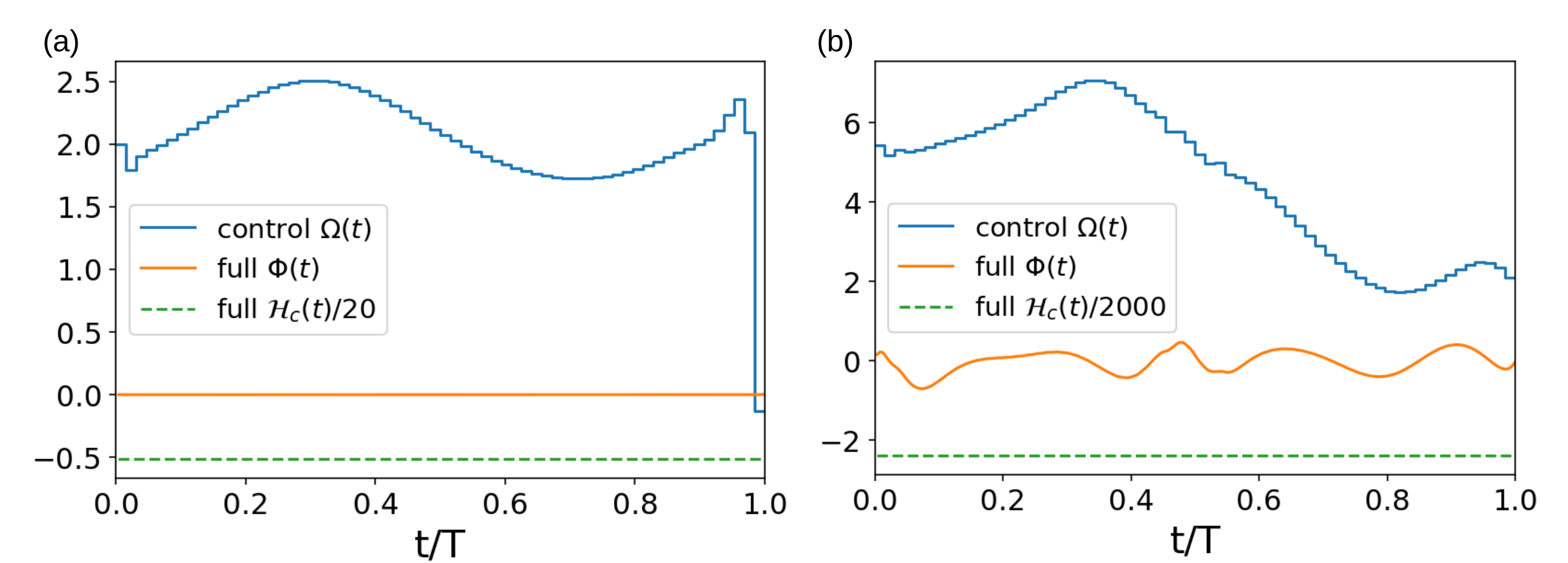}
\caption{(a) Optimal control for $N=4$, $\chi=1$ using 64 controls to maximize CFI.  The optimal CFI is about 8.19 with $\phi = \pi/2$.  (b) Optimal control for $N=100$, $\chi=0.1$ using 64 controls to maximize CFI.  The optimal CFI is about 2867.5 with $\phi = 0$.
}
\label{fig:N4_chi1_CFI}
\end{figure}

Results of $N=4$, $\chi=1$ and $N=100$, $\chi=0.1$ are presented in Fig.~\ref{fig:N4_chi1_CFI}(a) and (b). 64 time intervals are used to approximate $\Omega(t)$.  For $N=4$, $\chi=1$ [Fig.~\ref{fig:N4_chi1_CFI}(a)], both the mean and standard deviation are smaller than $10^{-3}$. 
For $N=100$, $\chi=0.1$ [Fig.~\ref{fig:N4_chi1_CFI}(b)], $\Phi(t)$ does not converge to zero but its mean is close to zero. The mean and standard deviation of $\Phi$ are respectively $\sim 0.06$ and $0.37$.  Overall, all necessary conditions are approximately satisfied for CFI optimization. 
As discussed in Refs.~\cite{PhysRevLett.119.193601, PhysRevLett.124.060402}, the measurement uncertainty can be taken into account by replacing $P_m$ by $\tilde{P}_m = \sum_{m'} \Gamma_{m,m'} P_{m'}$ in Eq.~\eqref{eqn:classical_FI} (with $\sum_m \Gamma_{m,m'} = 1$ for all $m'$). The proposed method can apply to this problem as well (not shown).

\section{Conclusion} 
To conclude, we apply Pontryagin's Maximum Principle to the quantum parameter estimation in the context of the ``twist and turn'' Hamiltonian. 
What PMP provides are (i) a formalism to efficiently evaluate the gradient with respect to a terminal cost function (the switching function); and (ii) a set of necessary conditions that can be used to quantify the quality of an approximate solution.
For the quantum metrology application, the performance is characterized by a single scalar -- the quantum or classical Fisher information, and the optimal control finds the  control protocol that maximizes QFI or CFI for a given evolution time.
One non-trivial complication pertaining to quantum metrology is that the cost function involves derivatives with respect to the external parameter which one wants to estimate, and we overcome this obstacle by designing an augmented dynamical system \ch{ where the wave function and its derivative to the external parameter $| \psi \rangle$ and $| \partial_\omega \psi \rangle$ are regarded as independent dynamical variables. 
By introducing the corresponding costate variables, all PMP quantities, particularly the switching function, can be stably obtained. }
The ability to efficiently compute the gradient greatly accelerates the optimization process and significantly expands the scope of problems one can solve. With the developed formalism, we are able to maximize QFI/CFI with more than 100 control variables. Moreover, the quality of an obtained control can be quantified by how well the PMP necessary conditions are satisfied (this applies to any approximate optimal controls). 
As a concrete example, we  show how an optimal solution converges upon increasing the number of controls by correlating the QFI and the smallness of the switching function. 
Specific to the ``twist and turn'' problem, we explicitly confirm the ``traditional sensing intuition'' in the strong twist limit: the main function of the optimal control is to steer the state to be close to $| \psi_\text{HL} \rangle$ (the state that maximizes QFI) quickly and then let the system freely interact with the environment. 
An important and natural question is the effect of quantum decoherence, and a quantitative answer requires calculations using density matrix as dynamical variables with dynamics involving dissipation channel(s). We expect the maximum QFI to occur at a finite evolution time (as a compromise between QFI $\sim t^2$ and decoherence), but the actual behavior should depend critically on the dissipation channel, especially when there is decoherence-free subspace \cite{PhysRevA.101.022320}. 

\section*{Acknowledgment}
We thank Yebin Wang (Mitsubishi Electric Research Laboratories) for very helpful discussions.
D.S. acknowledges support from the FWO as post-doctoral fellow of the Research Foundation -- Flanders. 
\bibliography{parameter_estimation}
\end{document}